# Test study on the RPS of TMSR-SF1 reactor *

LIU Zhenbao(刘珍宝)[1,2]　　LIU Ye(刘烨)[1]　　LIU Guimin(刘桂民)[1, †]
HOU Jie(后接)[1, ‡]

[1]Shanghai Institute of Applied Physics, Chinese Academy of Sciences, Shanghai 201800, China
[2]University of Chinese Academy of Sciences, Beijing, 100049, China

**Abstract:** The reactor protection system (RPS), as a 1E-level safety system, should be designed and developed following a series of nuclear laws and technical disciplines. The test, one process of the most rigorous requirement for the verification and validation (V&V), is the most significant and time-consuming effort and requires to test all functions by external response besides the review and analysis of the source code of the safety software. In this paper，a test system for the full digital and Field-Programmable Gate Arrays (FPGA) based RPS of the thorium based molten salt (TMSR) pebble bed fluoride salt-cooled reactor, so called solid fuel reactor (SF) is designed and applied to make sure the RPS fully meets its designed functions and system specifications. The purpose of the testing in this paper is to demonstrate whether RPS during the development is performed correctly, accurately and stability. We first introduce the principle and method of the test. Then the test hardware structures of the test system are designed and the software program is also developed. Finally the test process and the test results are discussed and summarized.

## Ⅰ INTRODUCTION

With the development of digital technology, Field-Programmable Gate Arrays (FPGAs) is gaining increased attention worldwide for safety I&C application in Nuclear Power Plants (NPPs) especially for reactor protection system (RPS) [1]. Compared to the conventional microprocessor software based technique, FPGA can improve the reliability and reduce the complexity for its certain gate circuit logic-based hardware feature [2]. FPGA has been applied in NPPs in a number of countries, such as USA, Ukraine [3]. Although FPGA technology is mature enough, FPGA based RPS for NPPs is still very new in the nuclear power industry. And as a 1E-level safety system, existing standard such as, NUREG/CR-7006 "Review Guidelines for FPGAs in NPP Safety Systems" demands that it should undergo more Verification &Validation (V&V) [4]. The V&V process demonstrates whether all stages during the development are performed correctly, completely, accurately, and consistently. Testing is one of the most significant and time-consuming effort in V&V, and requires to test all functions by external response besides the review and analysis of the source code of the safety software. However, because signal processing is done by FPGA, it is difficult to test the RPS status directly, as it is possible in analog systems by watching the relay operations. FPGA halt leads to the loss of signal processing capability due to software or hardware failure. Detection of failures in FPGA design and development may be more difficult and complex. The performance of the RPS, such as the response time of the RPS and the trigger precision, must also be determined by testing [5]. The lack of the development of more dedicated system and tools is the biggest challenge for V&V activity for FPGA-based systems.

* Supported by Leading Science And Technology Project of CAS (No. XD02001003).
†Corresponding author, liuguimin@sinap.ac.cn
‡Corresponding author, houjie@sinap.ac.cn

In fact, testing technology based on computer has been used in nuclear reactor engineering, for example, safety analysis and accident release measurement research [6], nuclear I&C design and V&V [7]. In the world, there are a few testing vendors such as CAE [8], WSC [9], who can provide large scale and full-scope NPP test system which price is very high. As a full digital FPGA based RPS of the thorium based molten salt (TMSR) pebble bed fluoride salt-cooled reactor, so called solid fuel reactor (SF), the structure and design principle is different from conventional NPP ones, which need specialized test system rather than large scale and full-scope one. Therefore a small scale and scope test system for the TMSR RPS is especially developed, by which all protection variables can be simulated in full range. And many design basis accidents of RPS, different types of failures can be tested and verified. Furthermore, an intuitive and friendly graphics user interface (GUI) makes the test system more humane in improving the interaction between the test system and human. This test system also greatly improves and speeds up the process of the V&V.

## Ⅱ. METHODS

### A. Principle of test

Generally, a test system generates test signals in a certain order and adds the test signals into the input interface of device. At the output interface of the device and some intermediate point, test system collects the test signals response after a certain time delay. Then the collected result is compared with the expected value to identify the consistency. If the two signals does not match well, it reveals that there are some abnormal states within the device which need to be checked out.

The RPS of TMSR-SF1 reactor is designed with a triple-modular redundant architecture as shown in Fig. 1. Each channel of the input modules receives the protection variables and passes the information to its respective FPGA processor. The three FPGA processors synchronize and communicate with each other by RS-485 protocol communication. The FPGA processors execute the logic program and send the results to the output modules. The output protection signal results from the 2-out-of-3 voting logic of the triggering signals. The surveillance station collects information of the RPS through RS-485. The system requirements are described summary in table 1.

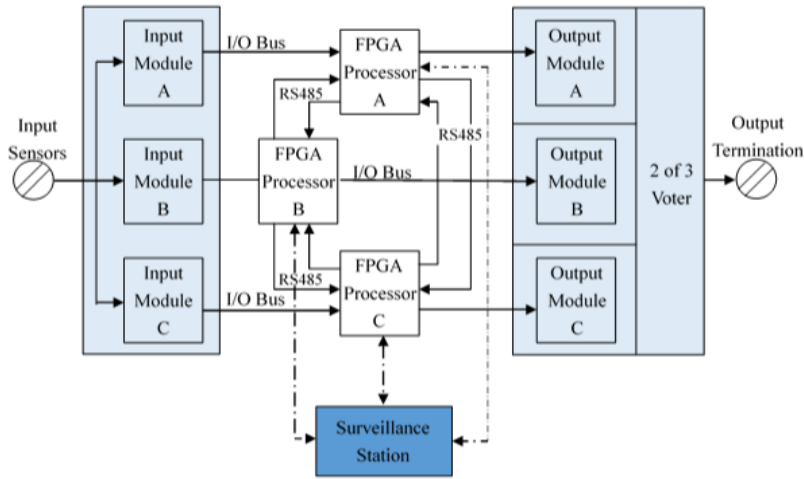

Fig. 1. Structure of RPS

Table 1. Summary of system requirements

| Category | Contents |
| --- | --- |
| I/O capacity | Meet the capacity requirements of Technical Specification |
| Response time | Less than 100ms |
| System reliability | Test time exceeds 1 month |
| Trigger set point | Less than 0.5 percent |
| System logic | Correctness |
| System failure | Correctness |

The test of RPS is divided into a number of "test sequences" which are executed by functionality in this study. Every test sequence corresponds to a protection function. Each sequence is also divided into several test steps. Each test step corresponds to a combination of input of the protection function. Test system sends a test signal to several input of the RPS synchronous in one test step, which can form different kinds of logic configurations. In order to ensure that the RPS has property of real signal response ability, a certain time delay is designed between two test steps. And the entity scope of the testing is from the input interface of the RPS to the output interface of the RPS.

### B. Structure of test system

The test system is a mobile automatic test equipment which will be connected with the RPS by test cable. Once the test is completed, the test system is broken away from RPS immediately. Fig. 2 illustrates the hardware block diagram of the test system which is composed primarily of power supply module, operator station (PXI platform-Slot PXI chassis and monitor terminal), five data acquisition cards (DAQs) of National Instrument Company (NI) and five signal conditioning boards (SCBs).

The test system can work for a long time due to its hardware based on PXI platform with powerful functions and high reliability which has higher precision and better stability than general control hardware. And it can also adapt to the environment of the reactor. With the help of PC client, edited program can install to the internal of PXI automatically without user intervention. PC monitoring program generated by LabVIEW runs on the PXI platform. The test program provides the human-machine interface (HMI), edits the test cases, manages the configuration files, transforms the program, analyzes and displays the test result, and controls the whole test system by HMI device-keyboard, mouse and displayer.

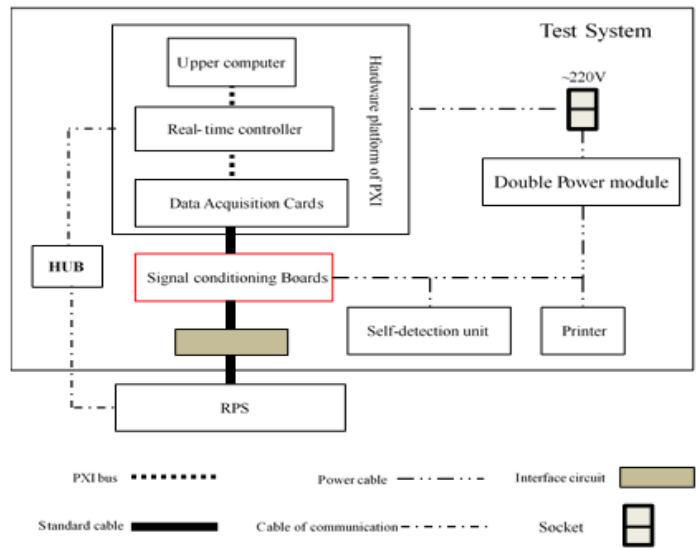

Fig. 2 Hardware Block Diagram of Test system

Conditioning, collection, input and output of analog and digital signals are realized by DAQ and SCB. The I/O signals of RPS is industry standard, such as standard 4-20 mA direct current and 0-24V direct voltage, similar to the NPP field signals. Considering the universality and compatibility, SCB is developed as the DAQ signals is not correspond to the RPS. This board is able to transform the DAQ signal to the type which the RPS required. To facilitate the operator to check the real-time SCB working condition, failure indication modular is designed in the board. SCB and DAQ are connected by standard cable, meanwhile the SCB and RPS are joined by an aviation connector and isolated by opt coupler.

In addition, the test system also achieve a self-detection function via a self-detection unit. This might ensure the correction of the system itself before the testing. The compare result of the process variable transmitters with setpoint values and logic conforms data of the RPS transfer through the serial communication way by a HUB port. Therefore, the test system is set in a standard cabinet to be safe and easy. The system also consider the expansion of the device interface in both circuit board and physical design.

### C. Work procedure of test system

The work procedure of test system is, during the working procedure of the test system, DAQ and SCB receive test output and control signals, all of which are from the computer. With the help of DAQ and SCB, the status signal which the protection logic circuit response to test injection signal, will be isolated and translated into test sampling signal which can be identified by a computer, then sent into the computer for reprocessing.

However, all of the protection variables can't be tested meanwhile, but tested one by one. For example, in the single test of low power range fixed value high neutron flux, only low power range fixed value high neutron flux disturbance is introduced, and other variables are kept stable. Associated variable will not be tested randomly, but be tested including linear load change by 5%/min and step load change by 10%.

### D. Software program and GUI

Software program, as an important factor in the test system, which should be the best selection of widely used in the world. Besides, many corresponding I/O models must be created to take into account cooperation with RPS. At present, the LabVIEW software [10] and the hardware of NI are applied in NPPs to transfer signals. So LabVIEW measurement and control software of NI Company are chosen. LabVIEW based on graphical G language has a multi-process concurrent execution and collects real-time characteristics of data, so that it is convenient to collect RPS's shutdown logic data in real-time. The software structure of test system is shown in Fig. 3, which is mainly divided into two parts of upper machine and lower machine. The order is stored and parsed in the upper machine when entered by an operator. After that, it is transmitted to the acquisition module and analysis module of the lower machine. Data acquisition and signal generation is completed by acquisition module. Data storage, data analysis and real-time response to operation instruction are all completed by the analysis module. Data collected from RPS is recorded and output a text file regularly.

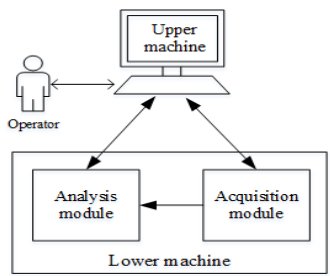

Fig. 3 Software architecture of test system

Software algorithm is designed to producer/consumer model as shown in Fig. 4. The software algorithm is that the lower machine receives data or order from upper machine in the first while cycle, then performs data acquisition or generates signal according to the instruction execution, and finally stores the data in the queue waiting for processing. Meanwhile, in the second while cycle it does series of operations to the data successively such as remove, analysis, store and display. Due to the elements in the queue is "first in and first out", the received data is ordered.

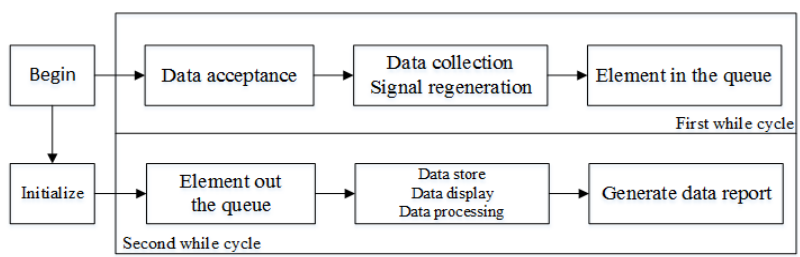

Fig. 4 Software algorithm model

In this test system, all monitoring and controlling works can be supported in the graphical user interface (GUI). GUI is used to monitor test process and control parameters, for example, manual/automatic control switch, set point adjustment, configuration changes, etc. According to the requirements, additional special functions are provided in GUI, such as signal process, data display, etc. The testing GUI of low power range fixed value high neutron flux as shown in Fig. 5. In addition, the physical configuration of I/O channels and its check are also performed in GUI.

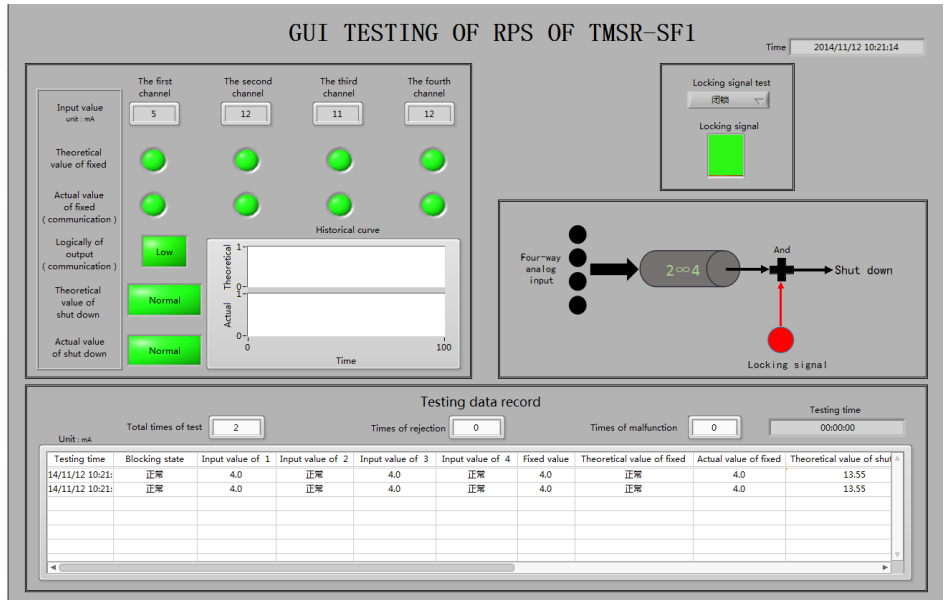

Fig. 5. GUI of low power range fixed value high neutron flux.

## Ⅲ. TESTING RESULTS AND DISCUSSION

The testing of the test system are selected and done in full accordance with GB/T 5204-2008 "Periodic tests and monitoring of the safety system of nuclear power plant".

### A. Reliability demonstration

As a testing defined in V&V plan and GB/T 5204-2008, and as the mandatory testing item required and witnessed by the research team, a long period reliability demonstration testing on the RPS has been performed. Within one month, the RPS has been running continuously, the protection signal is triggered by the test system repeatedly, the correctness of the triggering and the stability of triggering precision was recorded and verified by the test system, the number of triggering actions and then the number of overturns of output modules reaches $10^6$ times. The demonstration is more convincing and credible than analysis performed with the aim of proving the reliability of a system. This testing has proven and demonstrated the reliability of the RPS.

### B. Function testing

Besides the white box testing to the source code of safety software, such as review,

audit, analysis, unit test, simulation running etc., black-box testing (function testing) is performed on the RPS in the lab. This function covers all protection variables and their combination as well as the combination of them with the aid of the simulated signals from the test system. The aim of the function testing is to prove the following:

(a) Whether the sampling of signals and calculation of the protection variables are correct.

(b) Whether the RPS can cope with the full range of all input signals.

(c) Whether all protection logics are implemented correctly and the precision of trigger set point is within the required range.

(d) Whether all interlock logics and fail-safe mechanism to internal or external failures, such as the opening of a circuit, the cutting of a circuit, and a failure of any components or processors.

All of above are implement properly. This is one part of testing required for factory acceptance testing and site acceptance testing as defined in IEEE Std 7-4.3.2.

### C. Performance testing

The performance of the precision of the trigger set point, the stability of the sampling modules, the response time of the RPS, etc. have been tested.

Test results show that the precision of the trigger set point for each protection variables is within the designed range, and the stability of the sampling modules is very high. This proves the advantages of high precision and high stability in the information processing and transferring process in the RPS.

The response time of the RPS is measured by the test system. Protection variables of analog and digital are both chosen to trigger the RPS repeatedly and periodically. The test system records the input signal to the RPS and the state of the protection signal. The response time is determined by comparing between the time when the variable rose and the time when the state of output module output changed signal. For a digital device of RPS, each execution cycle will last a period which is not constant because of the communication between sampling modules and the RPS and between different parts of the RPS, and all functions including sampling, calculation, comparison with set point, trigger logic, output of protection signals and etc., will be executed only once in each cycle. After the signals of protection variables and the triggered status are recorded for a long time, for example 100 times, the maximum and the minimum value of the response time can be determined which are shown in Table 2.

Table 2. Results of response time of analog and digital

| Response time | Analog (ms) | Digital (ms) |
| --- | --- | --- |
| Maximum | 7.19 | 0.56 |
| Minimum | 1.01 | 0.22 |
| Average | 3.86 | 0.41 |

According to statistical theory, this response time is in accordance with the rule of normal distribution. So the record data is calculated to normal distribution fitting measurement in MATLAB software. Data histogram and the normal distribution curve fitting are got which are shown in Fig. 6. According to the probability density function,

$$f(x) = \frac{1}{\sqrt{2\pi}\sigma} e^{[-\frac{(x-\mu)^2}{2\sigma^2}]} \tag{1}$$

analog time: μ = 3.86, σ = 1.27, digital time μ = 0.41, σ = 0.06.

The analog response time within the credibility range of 95% is 3.61 ms to 4.11 ms, and the digital response time within the credibility range of 95% is 0.39 ms to 0.42 ms.

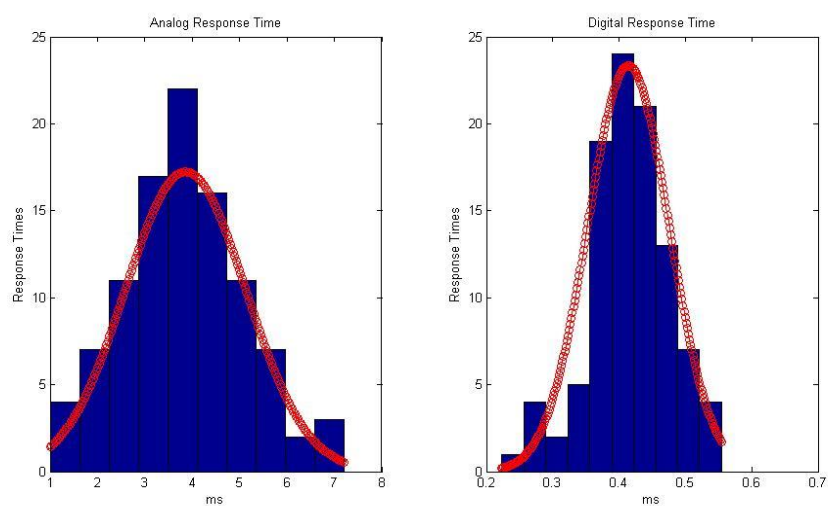

Fig. 6.  Data histogram and the normal distribution curve fitting of response time

### D. V&V of the test system

Guidelines or standards such as IEEE7-4.3.2 (1993), IEC880 (1986), JEAG4609 (1989) address the issues on reliability of safety systems and define requirements for hardware/software design, manufacturing, verification and validation (V&V) procedures documentation, maintenance and so on. The test system, as a non-nuclear safety system, is also undergo the V&V according above various regulatory standards and guidelines to judge whether the test system responds correctly. The procedure of V&V is well organized as shown in Fig .8. According to the procedure of V&V defined by the standards and/or guidelines, the following steps were taken.

Step 1: Quality assurance and software configuration are established according to the national and international standards. According to the standards requirement, a V&V team was set up, which was independent and responsible for all V&V activities. At the beginning, there was an original version of requirements and specifications documents in the stage of the test system development. A refined and ameliorated version documents were formed after the review of the original ones by the V&V team. And the updated documents were also studied and checked by all the engineers who developed the test system.

Step 2: The second step of V&V is to verify the hardware and software design specification, that is, the logic design specification. The logic design is verified by confirming that the logic described in the interlock block diagram meets the requirements of the system specification.

Step 3: The PXI and LabVIEW are widely used in many applications and fulfills the requirement of a test system, for example the environment adaptability of temperature, lifetime etc. For the verification of test system software-LabVIEW one part of the work is to review the design documentation and design scheme and to analyze and test the source code. These can be classified as "white-box testing". Another part of the work to analyze and test the software and the test system as a whole from the external response can be classified as "black-box test". The assessment of the source code of the test system software is done manually, and the conclusion is that the source code is verifiable, its behavior is deterministic, its performance is predictable, and the development process is reviewable.

Step 4: The system was tested independently using standard devices by V&V team as shown in Table 3. For example, a 4-20mA input in the test system by GUI will result in a same value at the output interface of the system which can be compared with analog signal through standard calibration instrument. Also the standard signal source simulation is test to calibrate the interval of two signals, the test result matches well which interval is 40 ms as shown in Fig. 7.

Table 3. The summary item of V&V Test

| Description | Data types | V&V result |
| --- | --- | --- |
| Output signals | 4-20mA | good |
| Receive signals | TTL、communication signals | good |
| Response time | 1~100ms | good |

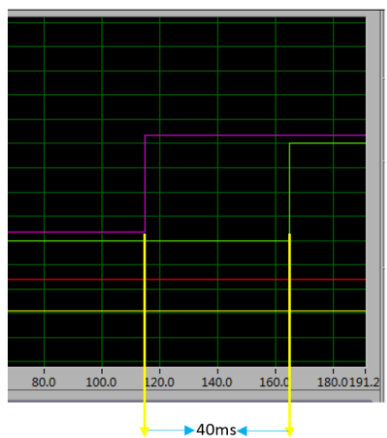

Fig .7 Test patterns of response time

The V&V results proves the advantages of high precision and high stability of the test system.

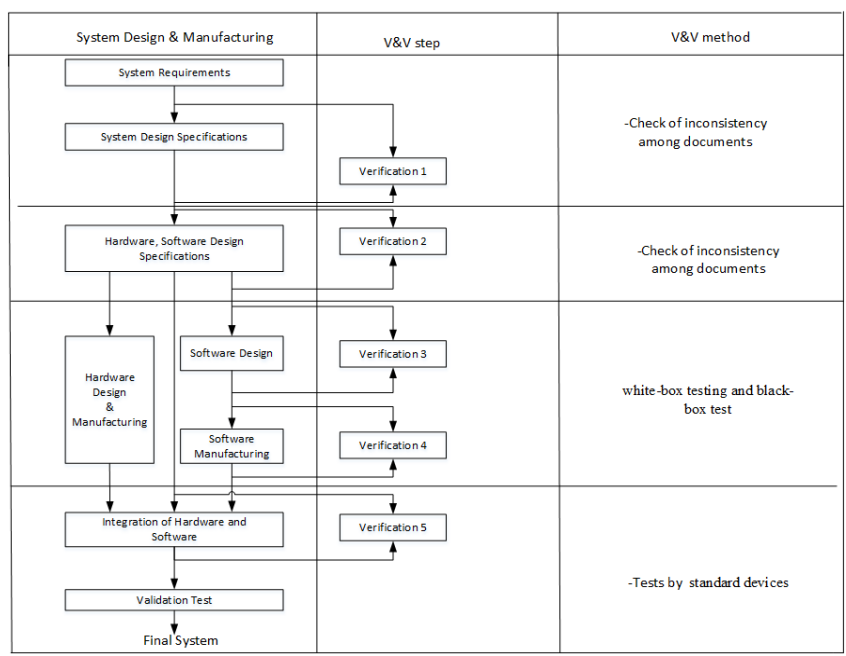

Fig. 8 Procedure of V&V

## Ⅳ. CONCLUSION

This test system can simulate all sensor parameters of the RPS to realize the test function, such as adding the simulation signals through the input/output interface equipment, collecting protection trigger signals, interning communication signals in real-time, and so on. All designs are aimed to accomplish the RPS's feature of high reliability, high redundancy and fast response time. Meanwhile the test system monitors the RPS's equipment status, tracks the location and estimates the faults once the RPS is out of order. I/O capacity of the test system can satisfy the requirement of test on RPS's performance. In addition, the test system has good performance with response time in milliseconds, analog signal output-precision in one over one thousand, and a long-term reliable and stably work.

The test system will be applied to $T_2$ and $T_3$ periodic experiments of protection system of TMSR-SF1 through an expansion. This test system has also greatly improved the efficiency and speeded up the process of the V&V, improve economy and keep safety during the development of RPS.

## ACKNOWLEDGMENTS

Supported by Leading Science And Technology Project of CAS (No. XD02001003). Authors are grateful for the support from all the people who were involved in this research.